\documentclass[pdflatex,sn-mathphys-num]{sn-jnl}

\usepackage{graphicx}%
\usepackage{multirow}%
\usepackage{amsmath,amssymb,amsfonts}%
\usepackage{amsthm}%
\usepackage{mathrsfs}%
\usepackage[title]{appendix}%
\usepackage{xcolor}%
\usepackage{textcomp}%
\usepackage{manyfoot}%
\usepackage{booktabs}%
\usepackage{algorithm}%
\usepackage{algorithmicx}%
\usepackage{algpseudocode}%
\usepackage{listings}%
\usepackage{tabularx}%
\usepackage[figuresright]{rotating}%
\usepackage{makecell}%
\usepackage{enumitem}%

\theoremstyle{thmstyleone}%

\theoremstyle{thmstyletwo}%

\theoremstyle{thmstylethree}%

\begin{document}

\title[Uncertainty in Atrial Fibrillation Detection]{Effect of Photoplethysmogram Artifacts on Uncertainty in Atrial Fibrillation Detection}

\author*[1]{\fnm{Andrius} \sur{Solo\v{s}enko}}\email{andrius.solosenko@ktu.lt}
\author[2]{\fnm{Enik\"o} \sur{Vargova}}\email{211218@vut.cz}
\author[3]{\fnm{Lo\"ic} \sur{Coquelin}}\email{loic.coquelin@lne.fr}
\author[1,4]{\fnm{Vaidotas} \sur{Marozas}}\email{vaidotas.marozas@ktu.lt}
\author[1,4]{\fnm{Andrius} \sur{Petr{\.{e}}nas}}\email{andrius.petrenas@ktu.lt}

\affil*[1]{\orgname{Biomedical Engineering Institute, Kaunas University of Technology}, \orgaddress{\street{Bar\v{s}ausko str. 59}, \city{Kaunas}, \postcode{51423}, \country{Lithuania}}}

\affil[2]{\orgname{Department of Biomedical Engineering, Brno University of Technology}, \orgaddress{\city{Brno}, \postcode{61600}, \country{Czech Republic}}}

\affil[3]{\orgname{National Laboratory of Metrology and Testing}, \orgaddress{\city{Paris}, \postcode{75015}, \country{France}}}

\affil[4]{\orgname{Department of Electronics Engineering, Kaunas University of Technology}, \orgaddress{\city{Kaunas}, \postcode{51368}, \country{Lithuania}}}

\abstract{Detection of atrial fibrillation (AF) from photoplethysmogram (PPG) is highly sensitive to artifacts, yet their effect on uncertainty of different AF detectors remains poorly understood. This work aims to quantify how different PPG artifact types affect the uncertainty of AF detectors. Two machine learning approaches to AF detection were explored: one using 25-s PPG signals as input ($\mathcal{D}_{r}$) and another using AF-related rhythm irregularity features ($\mathcal{D}_{f}$). The detectors were trained on wrist PPG signals acquired during cardiac rehabilitation and then systematically evaluated on 260,000 PPG signals containing controlled artifact types and durations. Uncertainty was quantified using a threshold-based error rate, conformal prediction, and Monte Carlo dropout. Using artifact-free PPG signals, $\mathcal{D}_{f}$ outperforms $\mathcal{D}_{r}$ with sensitivity/specificity of 0.94/0.94 versus 0.92/0.86. Relative to artifact-free performance, sensitivity/specificity drops by 0.52/0.03, 0.25/0.03, 0.16/0.02, and 0.07/0.02 using $\mathcal{D}_{r}$ for 12-s artifacts of device displacement, forearm motion, hand motion, and poor contact respectively. For $\mathcal{D}_{f}$, the respective drops are 0.31/0.07, 0.19/0.11, 0.14/0.15, and 0.13/0.21 for the same artifacts. $\mathcal{D}_{f}$ is more robust to short artifacts but exhibits increasing uncertainty with longer artifact durations, whereas $\mathcal{D}_{r}$ shows an abrupt performance drop when artifacts occur but is less sensitive to artifact duration. Applying conformal prediction with 90\% coverage reduces the false-positive rate by up to 12\% for $\mathcal{D}_{r}$ and up to 64\% for $\mathcal{D}_{f}$. Artifact type and duration have detector-specific effects on AF detection uncertainty. Device displacement causes the largest increase in uncertainty.}


\maketitle

\section{Introduction}
Detection of atrial fibrillation (AF) from photoplethysmogram (PPG) signals has gained popularity due to the widespread integration of PPG sensors into consumer wearables~\cite{pereira2020,papalamprakopoulou2024}. Large-scale studies using Apple~\cite{perez2019}, Huawei~\cite{guo2019mobile}, and Fitbit~\cite{lubitz2022detection} devices have shown that smartwatch-based PPG can be used to identify potential AF. However, these studies also reported many false positives, even when recordings were taken during periods expected to be free of movement. Although the sources of false positives were not analyzed in detail, premature beats and artifacts are known major contributing factors, as they can mimic AF-like pulse irregularity. While premature beats often exhibit repetitive rhythm interval patterns that can be suppressed algorithmically~\cite{solosenko2019detection}, artifacts remain much more challenging to address due to their diverse manifestations in PPG signals and their less understood effects on AF detection.

Two main approaches are commonly adopted for AF detection~\cite{pereira2020}. The first approach relies on engineered features that use clinical markers of AF, such as irregularity in pulse-to-pulse intervals, often supplemented by features that describe morphological or spectral properties of the PPG signal~\cite{millan2020analysis,avram2021validation,lubitz2022detection}. The second approach avoids manual feature engineering by training deep neural networks on preprocessed PPG signals or pulse-to-pulse intervals, for example, using a convolutional neural network (CNN)~\cite{Aschbacher2020,Kwon2020,antiperovitch2024}. Each approach has its own limitations: feature-based detectors depend heavily on accurate pulse peak detection and reliable feature extraction, whereas deep learning-based detectors may learn dataset-specific characteristics, including artifact-related ones, if these improve training performance.

An often overlooked aspect of AF detection is the quantification of uncertainty in the detector?s predictions. For a clinical screening tool, it is essential to assess how confident the detector is in each prediction, since low-confidence predictions are more likely to be incorrect and should not be relied upon. Most existing studies, however, report only aggregate performance measures such as accuracy or area under the receiver operating characteristic curve~\cite{pereira2020}, or simply classify borderline predictions as uncertain~\cite{antiperovitch2024}. To our knowledge, only one study has examined uncertainty in AF detection from PPG signals, using Monte Carlo dropout to estimate confidence in predictions of a deep learning-based detector~\cite{Bench2024towards}. That work, however, did not analyze how different artifact types influence uncertainty, nor did it explore feature-based detectors.

To address this gap, this study examines how PPG artifacts affect uncertainty in AF detection, focusing on artifacts due to device displacement, forearm motion, hand motion, and poor skin contact. Two approaches to AF detection are evaluated: a CNN that uses the PPG signal as input and a feature-based detector that relies on rhythm irregularity. By assessing uncertainty through a threshold-based error rate, conformal prediction, and Monte Carlo dropout, this study provides insight into AF detector behavior in the presence of artifacts and contributes to developing strategies for improving PPG-based AF screening.

\section{Materials}
\subsection{Training dataset}
The AF detectors, described in Sec.~\ref{SubDetectors}, were trained using a clinical signal database collected at the Kulautuva Rehabilitation Hospital, a branch of Kaunas Clinics of the Lithuanian University of Health Sciences. The database was acquired during cardiac rehabilitation following myocardial infarction, with patients being physically active during the observation period, including participation in guided workout sessions~\cite{paliakaite2021modeling}. The data acquisition protocol was approved by the Kaunas Region Biomedical Research Ethics Committee (No. BE-2-20).

Data were collected using a wrist-worn device developed at the Biomedical Engineering Institute of Kaunas University of Technology, Lithuania~\cite{paliakaite2017towards,solosenko2019detection}. The device synchronously acquires green-wavelength PPG at a sampling rate of 100~Hz, three-axis accelerations, and single-lead ECG signals. PPG signals were preprocessed using a zero-phase Butterworth bandpass filter with a passband of 0.5--5 Hz. Recordings were classified as AF or non-AF based on the reference ECG.

The AF dataset includes 15 patients with a mean age of \(72.9 \pm 8.9\) years, a body mass index of \(28.3 \pm 5.9\)~kg~m\(^{-2}\), and an observation period of \(21.3 \pm 3.8\)~h. The non-AF dataset consists of 19 patients with a mean age of \(67.5 \pm 10\) years, a body mass index of \(28.0 \pm 5.0\)~kg~m\(^{-2}\), and an observation period of \(21.6 \pm 3.1\)~h.

After balancing the dataset, the training set contains 37,102 AF and 37,102 non-AF signals, while the validation set contains 8,430 AF and 8,430 non-AF signals. No signals were excluded based on signal quality.

\subsection{Investigation dataset}
For detector investigation, the PPG simulator~\cite{solosenko2017modeling}, available through the open-access portal PhysioNet~\cite{solosenko2021model}, is used. The simulated PPG signal is obtained by positioning individual pulses according to the beat-to-beat intervals, thereby producing a continuous waveform in which each pulse is represented as a linear combination of one log-normal and two Gaussian functions. The realism of the simulated signals has been validated using synchronously acquired real signals, demonstrating that these closely replicate PPG morphology for normal rhythm, premature beats, and AF~\cite{solosenko2017modeling}. 

In this study, the simulator was further extended to include four types of artifacts commonly encountered in daily activities: device displacement, forearm motion, hand motion, and poor skin contact. Device displacement, when the sensor shifts or is jostled on the skin, causes the most pronounced distortion, often producing amplitude fluctuations up to an order of magnitude larger than normal PPG pulses~\cite{paliakaite2021modeling}. Forearm motion, such as swinging or shaking the arm, introduces slower baseline drift and background noise and is especially common during daytime activities. Hand motion, including actions like clenching fists or tapping fingers, produces abrupt changes in blood flow or minor sensor shifts, typically resulting in smaller but more transient artifacts. Poor sensor-to-skin contact, often due to a loose wristband, can cause intermittent signal loss.

Artifact extraction and parameterization are described in detail in~\cite{paliakaite2021modeling}. First, artifact-contaminated PPG segments were identified using the signal quality index~\cite{solosenko2019detection}. Then, artifacts were extracted by subtracting the modeled PPG signal from the acquired PPG signal. Finally, the extracted artifacts were classified into four types, each characterized by transition probability, duration, spectral slope, and normalized RMS amplitude.

Each type of artifact is produced by filtering white noise using a 250th-order finite impulse response filter, whose frequency response is defined by a spectral slope drawn from a Gaussian distribution with mean $\hat{\mu}$ and standard deviation $\hat{\sigma}$. The amplitude of the resulting artifact is scaled according to a normalized RMS amplitude taken from a gamma distribution with parameters $\hat{\alpha}$ and $\hat{\beta}$, representing the shape and scale of the distribution, respectively. Finally, the concatenated signal, comprising artifact and artifact-free intervals, padded with zeros, is added to the modeled pulsatile component to produce the simulated PPG signal. The parameters used for artifact simulation are summarized in Table~\ref{tab:artifact_parameters}.

\begin{table}[ht]
\caption{Parameters of spectral slope and normalized RMS for artifact types.}\label{tab:artifact_parameters}%
\begin{tabular}{@{}lccccc@{}}
\toprule
Artifact type & $\hat{\mu}_i$ & $\hat{\sigma}_i$ & $\hat{\alpha}_i$ & $\hat{\beta}_i$ & $\hat{\alpha}_i/\hat{\beta}_i$ \\
\midrule
Device displacement & -32.34 & 6.03 & 0.88 & 0.04 & 22.0 \\
Forearm motion & -29.39 & 5.71 & 1.44 & 0.31 & 4.65 \\
Hand motion & -25.45 & 4.13 & 1.40 & 0.45 & 3.11 \\
Poor contact & -18.12 & 4.10 & 2.01 & 1.24 & 1.62 \\
\botrule
\end{tabular}
\end{table}

The AF and non-AF PPG signals were simulated using beat-to-beat interval series extracted from ECG datasets: the MIT-BIH Normal Sinus Rhythm Database for non-AF and the Long-Term AF Database for AF~\cite{petrenas2017electrocardiogram}.

\begin{figure*}[h!]
\centering
\includegraphics[scale=0.75]{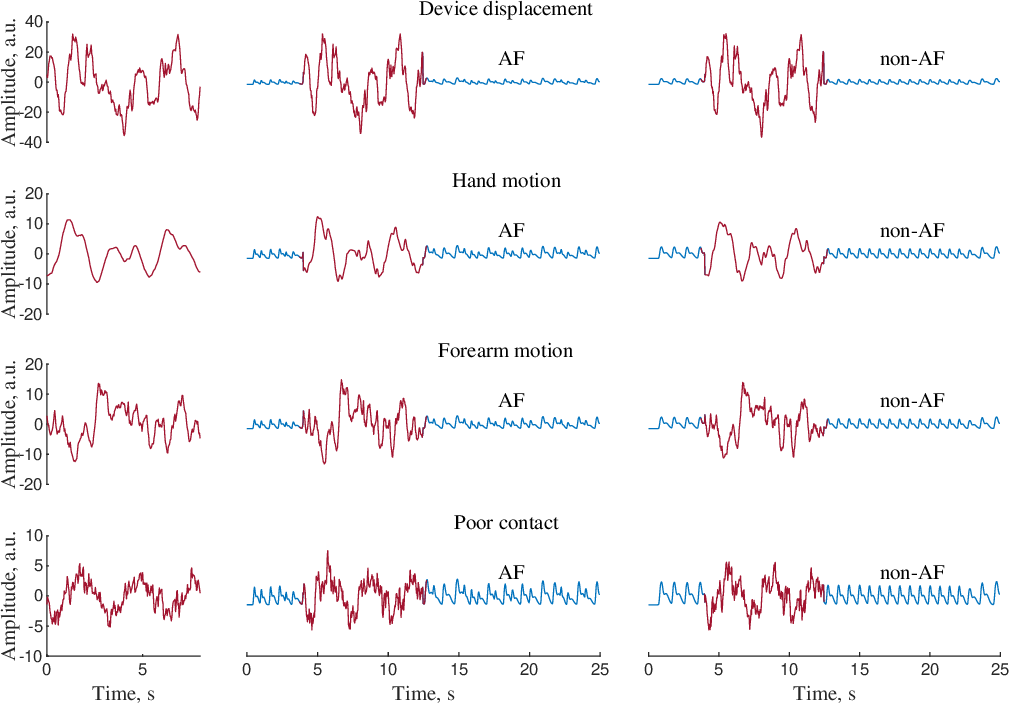}
\caption{Different types of simulated artifacts in the investigation dataset and their representation in the PPG signal. An added 8-second artifact is shown in red.}\label{Fig:Artifacts}
\end{figure*}

A set of 10,000 artifact-free simulated PPG signals with AF and 10,000 without AF, each 25 seconds long, were generated. For each artifact type, artifacts of 4, 8, and 12 seconds were added to this baseline set of 20,000 signals, referred to as $N$. In total, the dataset contains 20,000~+~20,000~$\times$~4~$\times$~3 = 260,000 PPG signals. For consistency, each artifact was inserted at the same time point within the simulated signal. An example of signals with different artifacts is shown in Fig.~\ref{Fig:Artifacts}.

\section{Results}
\subsection{Effect of artifacts on the prediction distribution}
\begin{figure*}[!ht]
\centering
\includegraphics[scale=0.75]{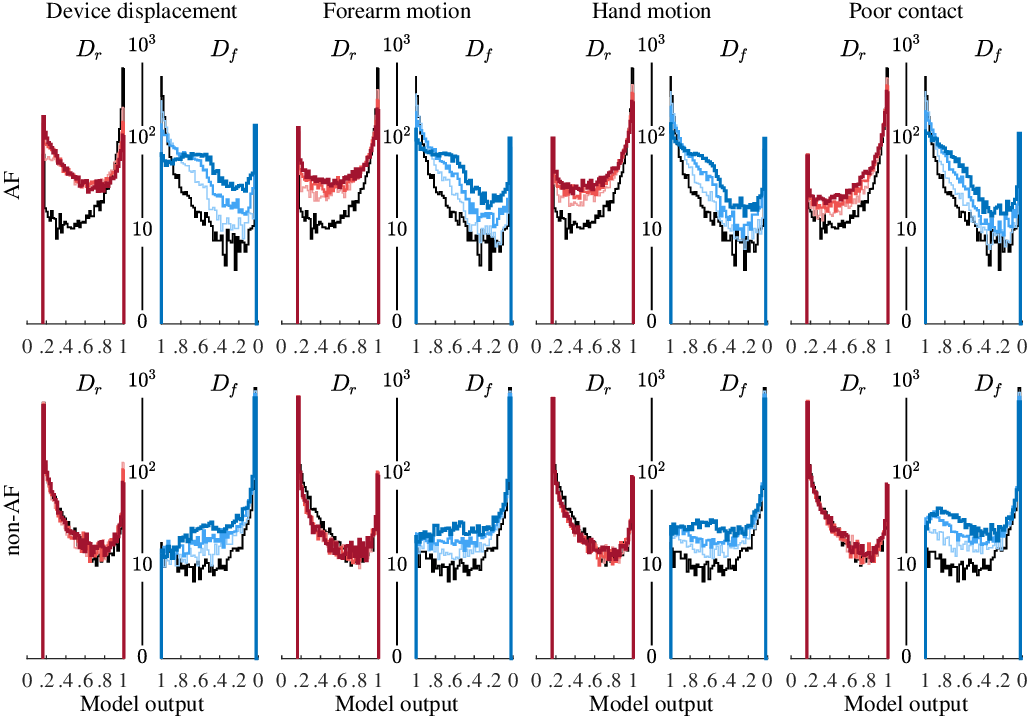}
\caption{Effect of artifacts on prediction frequency of detectors $\mathcal{D}_{r}$ and $\mathcal{D}_{f}$ for PPG signals with AF and non-AF. Black represents the distribution for signals without added artifacts, and progressively darker color shades indicate distributions for longer artifact durations, ranging from 4~s to 12~s in 4~s increments.}\label{Fig_histograms}
\end{figure*}

Figure~\ref{Fig_histograms} shows how each artifact type affects the prediction distributions of detectors $\mathcal{D}_{r}$ and $\mathcal{D}_{f}$. As expected, both AF and non-AF signals produce a bimodal distribution when the PPG is artifact-free. For $\mathcal{D}_{f}$, longer artifact duration flattens the distribution. In contrast, $\mathcal{D}_{r}$ reacts similarly to artifacts of any duration, with only slight flattening for non-AF and almost no change for AF. This may be due to the CNN treating artifacts and AF pulses as distinct from normal pulses. Across artifact types, the effect on the prediction distribution is similar, but device displacement has the strongest impact on $\mathcal{D}_{r}$ for non-AF signals. In that case, many predictions near 0.2 for artifact-free non-AF signals shift toward higher values.

\subsection{Effect of artifacts on detection performance}
\begin{figure}[!ht]
\centering
\includegraphics[scale=0.75]{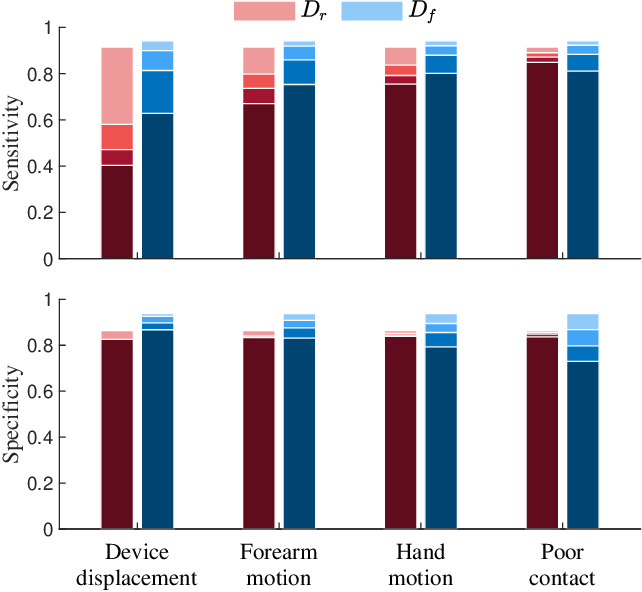}
\caption{Effect of artifacts on sensitivity and specificity of $\mathcal{D}_{r}$ and $\mathcal{D}_{f}$. Sensitivity is defined as the proportion of correctly detected AF cases, while specificity is the proportion of correctly detected non-AF cases. Shades ranging from lighter to darker indicate increasing artifact durations, starting from no artifact and progressing up to 12~s in 4~s increments.}\label{Fig_cnn_mlp_se_sp}
\end{figure}

Figure~\ref{Fig_cnn_mlp_se_sp} shows the effect of artifacts on the detection performance of $\mathcal{D}_{r}$ and $\mathcal{D}_{f}$ at a detection threshold of $\eta = 0.5$. In the absence of artifacts, $\mathcal{D}_{f}$ outperforms $\mathcal{D}_{r}$, achieving sensitivity/specificity of 0.94/0.94 compared to 0.92/0.86. Sensitivity decreases notably with longer artifact durations for both detectors, especially under device displacement. In contrast, the specificity of $\mathcal{D}_{r}$ is only minimally affected by artifact type and duration, whereas the specificity of $\mathcal{D}_{f}$ declines steadily as artifact duration increases. Relative to the artifact-free dataset, sensitivity/specificity drops by 0.52/0.03, 0.25/0.03, 0.16/0.02, and 0.07/0.02 for $\mathcal{D}_{r}$ under 12 s artifacts of device displacement, forearm motion, hand motion, and poor contact, respectively. For $\mathcal{D}_{f}$, the respective drops are 0.31/0.07, 0.19/0.11, 0.14/0.15, and 0.13/0.21.

\subsection{Threshold-based error rate}
\begin{figure}[!ht]
\centering
\includegraphics[scale=0.75]{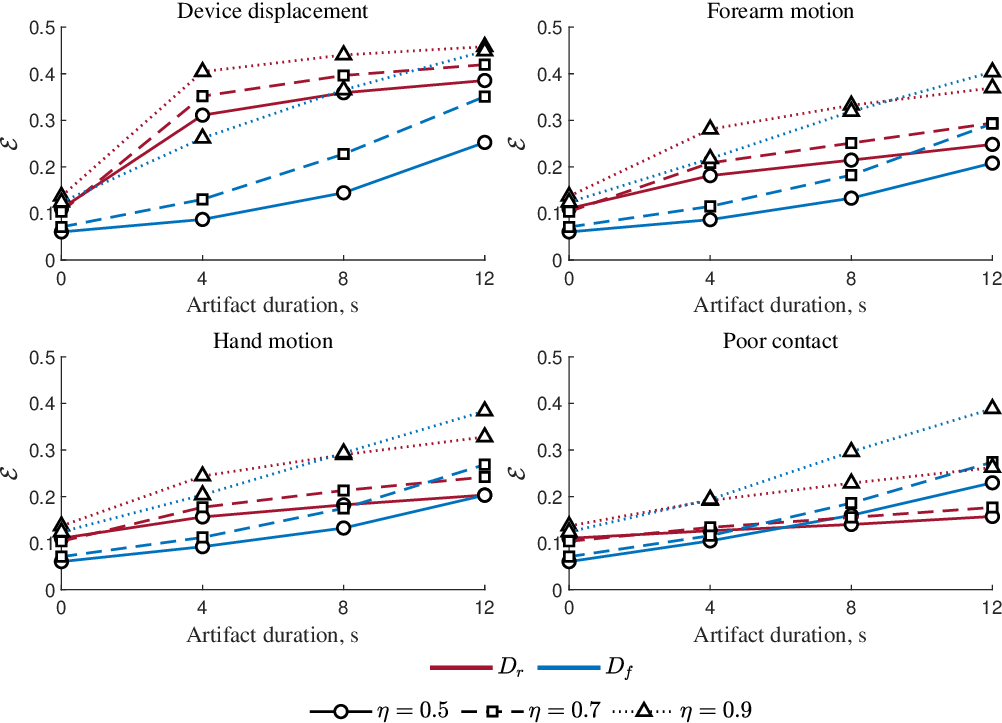}
\caption{Error rate $\mathcal{E}$ of decisions from $\mathcal{D}_{r}$ and $\mathcal{D}_{f}$ using different detection thresholds $\eta$.}\label{Fig_Error_rates}
\end{figure}

Figure~\ref{Fig_Error_rates} shows the error rates $\mathcal{E}$ of detectors $\mathcal{D}_{r}$ and $\mathcal{D}_{f}$ using detection thresholds of $\eta = 0.5$, $0.7$, and $0.9$. Increasing the threshold from 0.5 to 0.9 results in higher $\mathcal{E}$ for both detectors. The largest difference occurs under device displacement, where $\mathcal{E}$ of $\mathcal{D}_{r}$ is two to four times higher than that of $\mathcal{D}_{f}$ at an artifact duration of 4~s. In contrast, $\mathcal{D}_{r}$ generally yields lower $\mathcal{E}$ than $\mathcal{D}_{f}$ under poor contact. $\mathcal{D}_{r}$ exhibits higher $\mathcal{E}$ than $\mathcal{D}_{f}$ for shorter artifact durations; however, this difference decreases and eventually reverses in favor of $\mathcal{D}_{r}$ as artifact duration increases.

\subsection{Proportion of uncertain predictions}
\begin{figure}[!ht]
\centering
\includegraphics[scale=0.75]{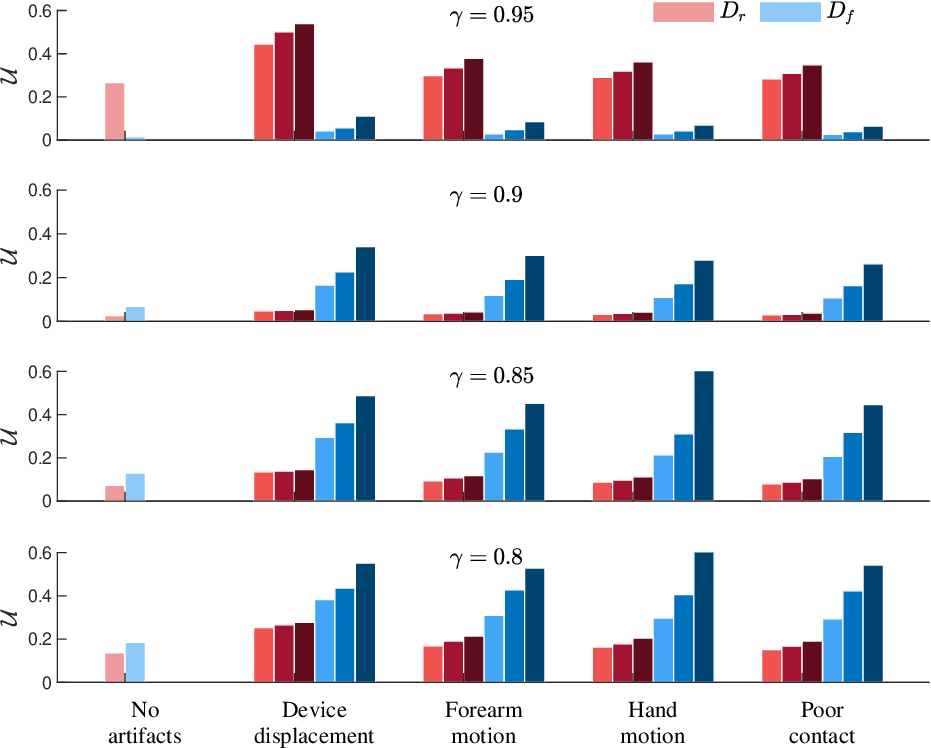}
\caption{Proportion of uncertain predictions $\mathcal{U}$ for different artifact types and durations for different coverage level $\gamma$. Shades ranging from lighter to darker indicate increasing artifact durations, starting from 4~s and progressing up to 12~s in 4~s increments.}\label{Fig_Uncertainty_proportion}
\end{figure}

Figure~\ref{Fig_Uncertainty_proportion} shows the proportion of uncertain predictions $\mathcal{U}$ produced by $\mathcal{D}_{r}$ and $\mathcal{D}_{f}$ using conformal prediction. At coverage levels $\gamma$ of 0.9, 0.85, and 0.8, $\mathcal{U}$ is consistently lower for $\mathcal{D}_{r}$ and nearly unaffected by artifact duration, while it steadily increases with artifact duration for $\mathcal{D}_{f}$. For example, at $\gamma = 0.9$, $\mathcal{U}$ is 0.06, 0.04, 0.04, and 0.04 for $\mathcal{D}_{r}$ under 12~s artifacts of device displacement, forearm motion, hand motion, and poor contact, respectively. For $\mathcal{D}_{f}$, the respective $\mathcal{U}$ is 0.34, 0.3, 0.28, and 0.26.

The opposite trend is observed for $\gamma = 0.95$, where $\mathcal{D}_{f}$ yields a much lower $\mathcal{U}$ than $\mathcal{D}_{r}$. This change in detector behavior is explained by the alignment of predictions with the conformal quantile: $\mathcal{D}_{f}$ produces predictions that span the full range $[0,1]$, while $\mathcal{D}_{r}$ output is restricted to $[0.16,1]$, see Sec.~\ref{SecDisc}.

Table~\ref{tab:conformal_prediction} sheds further light on how the prediction set composition changes with different values of $\gamma$. The output of $\mathcal{D}_{r}$ is constrained to a narrower range thus no non-AF decisions are produced for $\gamma = 0.99$. Since the detector is unable to produce very low predictions for non-AF PPG signals, the majority of them are assigned to the category ``both'', which increases $\mathcal{U}$.

\begin{table}[h]
\caption{Composition of the prediction set and proportion of uncertain decisions $\mathcal{U}$ at different coverage levels $\gamma$ for the artifact-free dataset.}\label{tab:conformal_prediction}%
\begin{tabular}{@{}lccccc c@{}}
\toprule
Detector & $\gamma$ & AF & non-AF & Both & Neither & $\mathcal{U}$ \\
\midrule
$\mathcal{D}_{r}$ & 0.99 & 2226 & 0    & 7774 & 0    & 0.78 \\
$\mathcal{D}_{r}$ & 0.95 & 4517 & 2818 & 2665 & 0    & 0.27 \\
$\mathcal{D}_{r}$ & 0.90 & 5131 & 4602 & 267  & 0    & 0.03 \\
$\mathcal{D}_{r}$ & 0.85 & 4941 & 4328 & 0    & 731  & 0.07 \\
$\mathcal{D}_{r}$ & 0.80 & 4762 & 3870 & 0    & 1368 & 0.14 \\
\midrule
$\mathcal{D}_{f}$ & 0.99 & 1605 & 4089 & 4306 & 0    & 0.43 \\
$\mathcal{D}_{f}$ & 0.95 & 4924 & 4908 & 168  & 0    & 0.02 \\
$\mathcal{D}_{f}$ & 0.90 & 4584 & 4725 & 0    & 691  & 0.07 \\
$\mathcal{D}_{f}$ & 0.85 & 4156 & 4549 & 0    & 1295 & 0.13 \\
$\mathcal{D}_{f}$ & 0.80 & 3720 & 4433 & 0    & 1847 & 0.19 \\
\botrule
\end{tabular}
\end{table}

\begin{figure}[!h]
\centering
\includegraphics[scale=0.75]{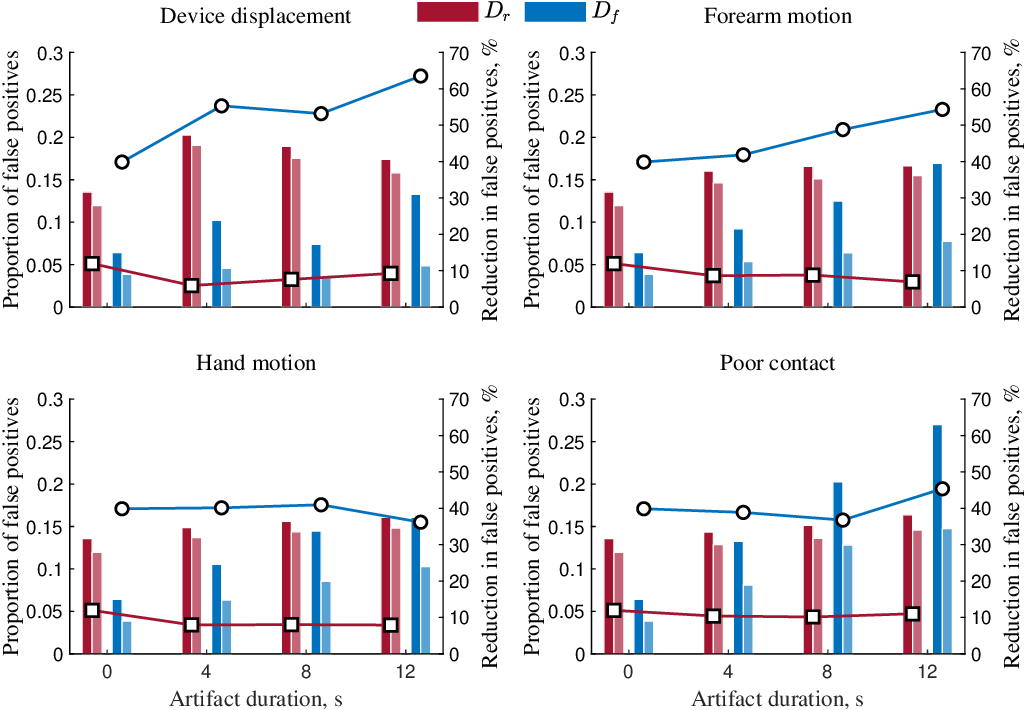}
\caption{Reduction in false positives (lines) obtained by discarding decisions flagged as uncertain using conformal prediction with $\gamma = 0.9$. Dark and light shaded bars represent the proportion of false positives before and after applying conformal prediction, respectively.}\label{Fig:FalseAlarms_misdetection}
\end{figure}

Figure~\ref{Fig:FalseAlarms_misdetection} shows the reduction in false positives by discarding decisions flagged as uncertain using conformal prediction with $\gamma = 0.9$. As the artifact duration increases, the proportion of false positives also increases. However, a growing number of these false positives are excluded by conformal prediction, resulting in reductions of false positives by 6--12\% for $\mathcal{D}_{r}$ and 36--64\% for $\mathcal{D}_{f}$ across different artifact types and durations.

\subsection{Effect of dropout}
\begin{figure*}[!ht]
\centering
\includegraphics[scale=0.75]{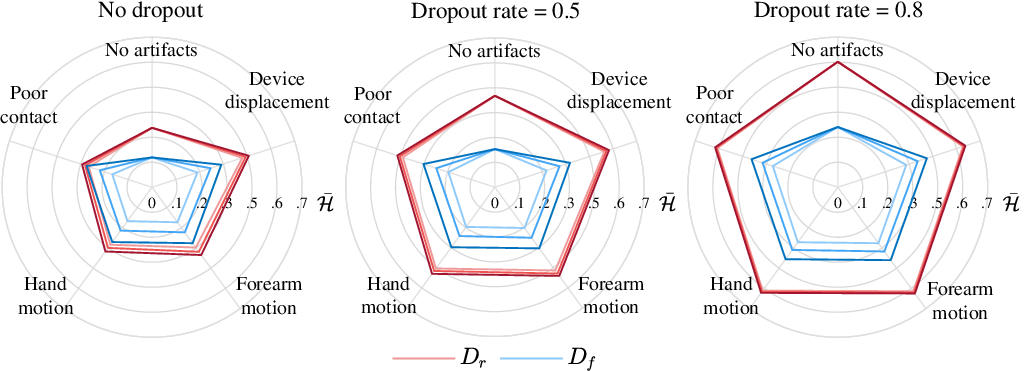}
\caption{Effect of dropout on the predictive entropy $\mathcal{\bar{H}}$ of $\mathcal{D}_{r}$ and $\mathcal{D}_{f}$. Shades ranging from lighter to darker indicate increasing artifact durations, starting from 4~s and progressing up to 12~s in 4~s increments.}\label{Fig_Predictive_Entropy}
\end{figure*}

Figure~\ref{Fig_Predictive_Entropy} shows the predictive entropy $\mathcal{\bar{H}}$ for predictions using Monte Carlo dropout with no dropout, and dropout rates of 0.5 and 0.8. When activations are zeroed during dropout, different sub-networks are created, processing the PPG signal differently from the fully connected network. This increases prediction variability, which in turn raises entropy $\bar{\mathcal{H}}$.

For both detectors, increasing the dropout rate leads to higher $\mathcal{\bar{H}}$, with the largest entropy observed at a dropout rate of 0.8 and the lowest with no dropout. However, the differences in $\mathcal{\bar{H}}$ across varying artifact durations are less pronounced in $\mathcal{D}_{r}$. This result is consistent with earlier findings: $\mathcal{D}_{r}$ is less affected by changes in artifact duration.

For all artifact types and durations, $\mathcal{\bar{H}}$ is lower for $\mathcal{D}_{f}$ compared to $\mathcal{D}_{r}$, suggesting that $\mathcal{D}_{r}$ relies on more interacting learned features. Since $\mathcal{D}_{f}$ operates on engineered, lower-dimensional features, its internal representations appear less sensitive to dropout. In fact, except for poor contact, $\mathcal{\bar{H}}$ for $\mathcal{D}_{f}$ in the presence of artifacts is still lower than $\mathcal{\bar{H}}$ for $\mathcal{D}_{r}$ without any artifacts.

\section{Discussion}\label{SecDisc}
The main finding of the study is that the feature-based detector $\mathcal{D}_{f}$ performs better when artifacts are short but becomes more uncertain as artifacts last longer. On the other hand, the detector $\mathcal{D}_{r}$, which operates on PPG signals, exhibits an abrupt drop in performance when artifacts occur but is less sensitive to artifact duration.

Differences in how the detectors handle artifacts may explain their diverging performance in noisy conditions. $\mathcal{D}_{f}$ uses the signal quality index as an input, which may contribute to the gradual increase in detection error for longer artifacts. Using PPG signals directly as input to a CNN is the conventional strategy~\cite{pereira2020,Kwon2020,bulut2025}; therefore, we did not consider architectures that incorporate the signal quality index. $\mathcal{D}_{r}$ has 800 inputs corresponding to the number of samples in each PPG signal thus adding a signal quality index value as another input is inappropriate due to the mismatch in dimensionality. An alternative is to let the CNN first extract features from the waveform and then combine the signal quality index with these features at a later stage, for example, in the fully connected layers. Such late-fusion methods have been shown to improve performance when combining deep features with signal analysis-based features~\cite{xiao2022deep}.

The finding that a feature-based detector can perform comparably to that trained on PPG signals aligns with the literature. For example, a long short-term memory network trained on peak-to-peak intervals performed similarly to a more complex CNN trained on PPG signals~\cite{isaksen2025}. Similarly, comparable performance was reported for a heuristic signal-processing-based detector and a CNN-based detector, although the latter excluded less data due to poor signal quality~\cite{antiperovitch2024}.

Uncertainty in detector predictions can arise from epistemic and aleatoric sources~\cite{hullermeier2021}, but the boundary between the two is often unclear~\cite{der2009aleatory}. For this reason, we did not attempt to separate them in this study. Epistemic uncertainty due to limitations in the detector architecture and insufficiently representative training data, can be reduced using a better machine learning model, training on a larger and more diverse dataset, handling artifacts more effectively, or incorporating more informative features~\cite{hullermeier2021}. Uncertainty related to variability in PPG signals is less well defined, as it depends on factors such as sensor design, PPG acquisition settings, and artifacts~\cite{charlton2025}. Part of this uncertainty is aleatoric, such as pulse-to-pulse fluctuations caused by natural physiological processes. Artifacts exhibit characteristics of both uncertainty types: they may behave like random processes, yet they can also be reduced using better sensors and signal processing algorithms.

The coverage level $\gamma$ in conformal prediction strongly influences the uncertainty estimates, and this effect depends on the range of output values produced by the detector. As shown in Fig.~\ref{Fig_histograms}, $\mathcal{D}_f$ predictions span the full range $[0,1]$, whereas predictions from $\mathcal{D}_r$ take a narrower interval. Consequently, at high coverage levels, e.g., $\gamma = 0.95$, $\mathcal{D}_r$ yields a much higher proportion of uncertain decisions $\mathcal{U}$ (see Table~\ref{tab:conformal_prediction}) because it cannot produce low prediction values for the non-AF class. These findings suggest that uncertainty assessment based solely on $\mathcal{U}$ can be misleading if $\gamma$ and the detector's output are not considered together, as detectors with narrower prediction ranges are penalized more severely.

Deep neural networks often exhibit overconfidence~\cite{guo2017calibration}, but $\mathcal{D}_r$ behaves differently, producing less confident predictions for the non-AF class. Changing CNN hyperparameters, for example, reducing the batch size or adding more convolutional layers, increases the prediction range but at the cost of reduced performance. Future work should investigate architectural modifications to enable wider output ranges, complemented by calibration methods such as temperature scaling~\cite{guo2017calibration}.

In conformal prediction, the calibration and test datasets are assumed to be exchangeable, i.e., drawn from the same distribution; this assumption does not formally apply to the training data~\cite{vovk1999machine,angelopoulos2021gentle}. However, a mismatch between the training and test data can still affect calibration~\cite{Tibshirani2019}. In our study, PPG signals used for training were collected from patients, whereas those used for calibration were artifact-free simulated. To handle distribution mismatch in practice, distribution-aware conformal methods can be used, such as weighted conformal prediction, which adjusts the influence of calibration samples based on how well they match the test distribution~\cite{Tibshirani2019,yang2024doubly}.

Artifacts were extracted from PPG signals collected in controlled settings and during cardiac rehabilitation using a wrist-worn device with a green-light sensor~\cite{paliakaite2021modeling}; the wavelength commonly used in smartwatches~\cite{charlton2023}. Consequently, the simulated artifacts approximate those typical in free-living conditions, particularly among older adults. It should be noted, however, that the occurrence of artifacts is known to vary over time and context, with a higher prevalence observed during daytime periods~\cite{paliakaite2021modeling}. This is primarily attributed to increased forearm motion and device displacement~\cite{pradhan2019evaluation,paliakaite2021modeling}.

Device displacement had the strongest negative impact on the performance of both detectors, especially for $\mathcal{D}_{r}$. This is likely due to the relatively large amplitude of this artifact. Therefore, future studies should carefully consider the choice of normalization technique; not only to account for inter-individual variability in heart rate and signal characteristics~\cite{gasparini2022}, but also for the presence of artifacts. For example, applying fixed-range normalization such as min-max scaling~\cite{ding2023,bulut2025} can cause large artifacts to dominate the scaling range, compressing the physiological portion of the signal and increasing the relative influence of artifacts on detector performance. Therefore, to reduce the impact of large-amplitude artifacts on scaling, we used z-score normalization instead of fixed-range scaling. However, other approaches, such as soft clipping or nonlinear amplitude transformations, e.g., passing the input PPG signal through a hyperbolic tangent function, could also be considered.

Conformal prediction communicates uncertainty in a straightforward way, which helps support clinical decision-making~\cite{kapuria2024,papangelou2025}. If both classes are returned for a given PPG signal, the detector cannot reliably distinguish between AF and non-AF, whereas an empty set indicates that the automatic analysis should not be relied on. Due to the lack of established guidelines for interpreting arrhythmias from PPG, verification is required either by collecting an intermittent ECG recording with the same device or by referring the patient for further clinical examination. If conformal prediction returns AF, the user should be prompted to confirm the suspected AF with ECG. If the result is both classes or an empty set, the device should suppress the alert and avoid unnecessary ECG acquisition or user notification. In this study, suppressing such uncertain outputs led to substantial reductions in false positives, reaching up to 12\% for $\mathcal{D}_{r}$ and up to 64\% for $\mathcal{D}_{f}$, see Fig.~\ref{Fig:FalseAlarms_misdetection}.

A practical way to exploit the complementary strengths of the two approaches to AF detection is to use both detectors in parallel and select the more certain output based on artifact duration. Our findings show that $\mathcal{D}_{f}$ outperforms $\mathcal{D}_{r}$ when analyzing PPG signals with short duration artifacts, whereas $\mathcal{D}_{r}$ becomes comparatively more certain as artifact duration increases. This supports the use of a gating strategy: for example, assess PPG signal quality together with the width of the conformal prediction set, and if $\mathcal{D}_{f}$ exhibits uncertainty and the signal contains a large proportion of artifacts, switch to $\mathcal{D}_{r}$. Since modern cloud platforms can run both detectors with minimal latency, such a dual-inference approach is feasible for real-time use. Similar hybrid frameworks have been proposed in the literature; for example, DeepAware combines a deep neural network with context-aware heuristics to reduce false positives in ambulatory AF detection~\cite{kumar2022deepaware}.

A limitation of the present study is that investigation relies on simulated PPG signals and artifacts. AF detection performance on artifact-free simulated signals is comparable to that obtained using real PPGs~\cite{Bashar2019,dorr2019,nonoguchi2022,selder2023,zhao2024}, though simulations may not fully cover the complexity of real-world data. The use of simulated signals and artifacts is motivated by the need to systematically control artifact characteristics. While real signals with added simulated artifacts can be an option, signals acquired from a wrist often contain residual artifacts. For example, nearly one-third of nighttime PPG signals from resting individuals have been reported as uninterpretable due to poor signal quality~\cite{valiaho2022continuous}. These residual artifacts can obscure the true impact of simulated artifacts on AF detection.

Simulation of artifacts mostly focuses on external factors such as motion and variations in sensor attachment. Consequently, it does not account for intrinsic PPG signal perturbations that originate from underlying physiological processes. These include respiration-induced baseline fluctuations, vasomotor tone changes related to thermoregulation, and transient vasoconstriction responses associated with stress or sympathetic activation~\cite{allen2007}. Such physiological sources of variability can modulate pulse amplitude and morphology independently of movement or contact quality and may therefore contribute to uncertainty in detector predictions.

The training dataset consisted of older cardiac rehabilitation patients. Younger or healthier individuals, different skin tones, alternative sensors, e.g. different wavelengths or positioning, might produce different signal and artifact characteristics.

\section{Conclusions}
This study shows that machine-learning approaches to AF detection respond differently to artifacts in PPG signals. The detector based on AF-related irregularity features outperforms the CNN under short artifacts but becomes increasingly uncertain as artifact duration increases. The CNN exhibits a substantial performance drop in the presence of artifacts, yet it is less affected by increasing artifact duration. Among the artifact types evaluated, device displacement has the most pronounced negative impact on detection performance.

\section{Methods}
\subsection{AF detection}\label{SubDetectors}
\subsubsection{Detection using PPG signals}
\begin{figure}[h!]
\centering
\includegraphics[scale=0.65]{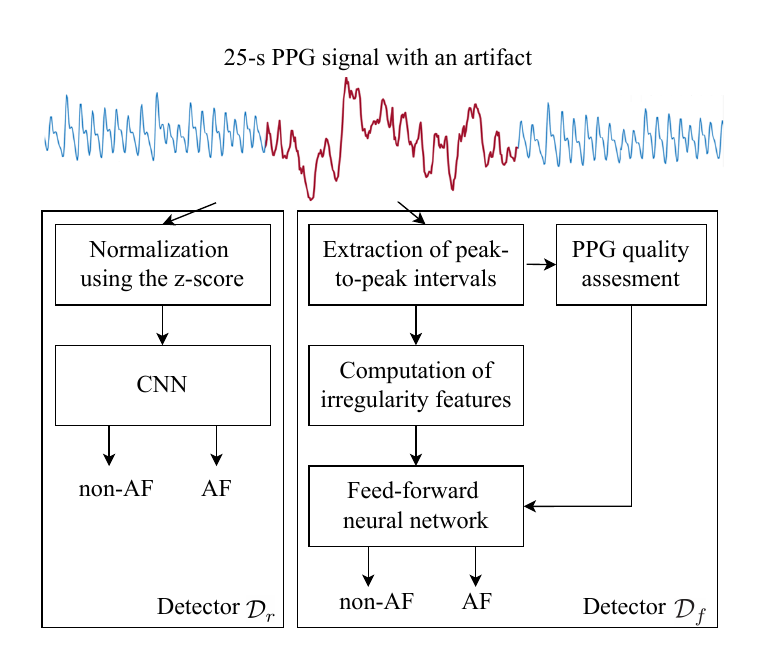}
\caption{Block diagrams of the AF detectors under investigation: detector $\mathcal{D}_{r}$ that uses the PPG signal as input, and detector $\mathcal{D}_{f}$ that uses AF-related rhythm irregularity features with a signal quality index.}\label{Fig:detectors}
\end{figure}

Figure~\ref{Fig:detectors} shows a block diagram of the CNN-based AF detector, denoted as $\mathcal{D}_{r}$, which follows the architecture described in~\cite{butkuviene2021}. The 1D~CNN takes as input a vector of 800 samples from a 25-second PPG signal sampled at 32~Hz, normalized using z-score standardization. The network consists of two convolutional layers with 128 and 64 kernels of size 64 and 32 samples, respectively. Each convolutional layer is followed by max pooling with a pool size of 2 and stride of 2, and then a fully connected layer with 128 neurons. All hidden layers use the rectified linear unit activation function and a dropout rate of 0.5 to reduce overfitting. The output layer consists of two units, activated by the softmax function, corresponding to the classes AF and non-AF. The detector is trained using the Adam optimizer~\cite{kingma2014adam}, with a mini-batch size of 512, learning rate of 0.001, and cross-entropy loss as the objective function. Training is performed for 1000 epochs, with early stopping applied if validation accuracy does not improve for 40 consecutive epochs.

\subsubsection{Detection using engineered features}
Figure~\ref{Fig:detectors} shows a block diagram of the detector, denoted as $\mathcal{D}_{f}$, which uses AF-related irregularity features and a signal quality index as input to a feed-forward neural network. The hidden layer contains 128 neurons, followed by a dropout layer with a rate of 0.5. The Adam optimizer is used to train the detector~\cite{kingma2014adam}, with a mini-batch size of 32, a learning rate of 0.01, and cross-entropy loss as the objective function. The rectified linear unit activation function is applied to the hidden layer neurons, while a softmax function is used to produce probability distributions.

The occurrence times of PPG pulses are identified using the peak detector described in~\cite{solosenko2019detection}, producing a sequence of peak-to-peak intervals. PPG signal quality is assessed using a signal quality index that compares each PPG pulse with a template~\cite{solosenko2019detection}. A correlation threshold of 0.7 is used to define acceptable quality. The signal quality index represents the proportion of usable signal relative to the total duration and is included as an additional feature.

The irregularity in peak-to-peak intervals, a characteristic of AF, is quantified using seven features that characterize randomness, variability, and complexity~\cite{sornmo2018atrial}.

The randomness is evaluated using the turning point ratio and Shannon entropy. Turning point ratio evaluates the randomness in peak-to-peak intervals by comparing each interval to its two nearest neighbours~\cite{dash2009automatic}. An interval is considered as a turning point if it is greater or less than its two neighbours. Then, the turning point ratio is computed as the proportion of turning points to the total number of intervals. A higher turning point ratio signifies a greater degree of randomness. Shannon entropy quantifies the unpredictability of peak-to-peak intervals~\cite{shannon1948mathematical} and tends to increase when the interval distribution is less concentrated around the specific value. Shannon entropy is typically higher during AF than during non-AF rhythm~\cite{zhou2014automatic}.

The variability is assessed with four features. The coefficient of variation relates the standard deviation of peak-to-peak intervals to their mean and tends to increase during AF, as variability rises and the mean interval shortens~\cite{tateno2001automatic}. The mean successive beat interval difference measures variability between consecutive intervals relative to the mean~\cite{langley2012accuracy}, while the root mean square of successive differences provides a related measure that is independent of mean heart rate~\cite{dash2009automatic}. In addition, the Poincar\'e plot visualizes consecutive peak-to-peak intervals as a scatter plot~\cite{park2009atrial,lian2011simple}. Plots during AF show much greater dispersion than those from non-AF rhythm. A fixed 56 by 56 grid of 25~ms bins is placed over the plot, covering peak to peak intervals from 200 to 1600~ms and peak to peak differences from -500 to 900~ms. For each window, the number of grid cells that contain at least one point is counted and used as the feature value.

The complexity is evaluated using sample entropy. Sample entropy assesses the self-similarity present within peak-to-peak intervals~\cite{richman2000physiological}. Sample entropy is expressed as the negative natural logarithm of the conditional probability that sequences of length \textit{m} that match within tolerance \textit{r} will also match for length \textit{m}+1, with self-matches excluded. In this study, \textit{m}~=~1 and \textit{r}~=~30~ms~\cite{petrenas2015low}.

\subsection{Uncertainty quantification}
Uncertainty in detector performance is quantified using three approaches: threshold-based error rate, conformal prediction, and Monte Carlo dropout.

\subsubsection{Threshold-based error rate}
To assess the detection error at different decision thresholds, the detector outputs are thresholded at $\eta = 0.5$, $\eta = 0.7$, and $\eta~=~0.9$ to produce AF or non-AF class decisions. Each predicted class $\hat{y}_i$ is compared with the true class $y_i$, and every mismatch is counted as an error. For each threshold, the detection error rate is computed as the total number of errors divided by the total number of PPG signals:
\begin{equation}
\mathcal{E}(\eta) = \frac{1}{N} \sum_{i=1}^{N} \mathbf{1}\!\left( \hat{y}_i(\eta) \neq y_i \right).
\end{equation}

\subsubsection{Conformal prediction}
Conformal prediction returns a decision set that may contain one class, both classes, or no class at all, rather than making a binary AF vs non-AF decision.

After training the detector on the training dataset, nonconformity scores are computed on the calibration set of size $C$. For a PPG signal $c_i$ belonging to the class $y_i$, the nonconformity score is defined as
\begin{equation}
\alpha_i =
\begin{cases}
1 - p(c_i), & y_i = \text{AF},\\
p(c_i), & y_i = \text{non-AF}.
\end{cases}
\end{equation}
where $p(c_i)\in[0,1]$ is the detector output for the AF class. A lower $\alpha_i$ indicates higher confidence in prediction.

Given the set of calibration scores $\{ \alpha_1, \alpha_2, \dots, \alpha_C \}$, sorted in ascending order as $\alpha_{(1)} \le \cdots \le \alpha_{(C)}$, the nonconformity threshold is obtained as
\begin{equation}
m = \left\lceil (C + 1)(1-\gamma)\right\rceil, \qquad \alpha_q = \alpha_{(m)},
\end{equation}
where $\alpha_{(m)}$ is the $m$-th smallest calibration score, $\lceil \cdot \rceil$ is the ceiling operator, and $\gamma$ is a predefined coverage level, which is the probability that the true class is included in the prediction set.

For each PPG signal $x_i$ of the investigation dataset, the nonconformity scores are computed as
\begin{equation}
\alpha_{\text{AF}}(x_i) = 1 - p(x_i), \qquad
\alpha_{\text{non-AF}}(x_i) = p(x_i),
\end{equation}
and the conformal prediction set is formed:
\begin{equation}
\Gamma(x_i) = \{\, y : \alpha_y(x_i) \le \alpha_q \,\}.
\end{equation}
$\Gamma(x_i)$ is interpreted as follows: if $\Gamma(x_i)=\{\text{AF}\}$, the signal is classified as AF; if $\Gamma(x_i)=\{\text{non-AF}\}$, it is classified as non-AF; if $\Gamma(x_i)=\{\text{AF},\text{non-AF}\}$, the prediction is considered uncertain; and if $\Gamma(x_i)=\varnothing$, the prediction is rejected due to insufficient confidence.

Empty sets occur when no class achieves a conformity score below the threshold, suggesting that the PPG signal differs substantially from the calibration distribution. Two-class sets occur when both classes have a conformity score below the threshold, indicating ambiguity between them. Although these outcomes arise for different reasons, we combine them into a single uncertainty measure defined as the proportion of cases in which the detector does not return a single-class decision:
\begin{equation}
\mathcal{U} = \frac{1}{N} \sum_{i=1}^{N} \mathbf{1}\{\,|\Gamma(x_i)| \neq 1\,\},
\end{equation}
where $\mathbf{1}\{\cdot\}$ is the indicator function that equals 1 when the decision set contains either both classes or neither class.

To compose a calibration dataset, artifact-free PPG signals were randomly split into two equally sized subsets. The calibration dataset consists of $C = N/2$ artifact-free PPG signals, while the remaining $N/2$ signals, together with all artifact-contaminated signals, are used for testing. Only artifact-free PPG signals are used for calibration to ensure that the empirical distribution of nonconformity scores reflects detector uncertainty rather than artifact-induced distortions, which would affect the conformal prediction threshold.

\subsubsection{Monte Carlo dropout}
Monte Carlo dropout is used to estimate uncertainty in deep neural networks by applying dropout, which serves as an approximation of Bayesian inference~\cite{gal2016dropout}. Dropout remains active during inference in both the convolutional and fully connected layers of $\mathcal{D}_r$ and in the hidden layer of $\mathcal{D}_f$, causing the predictions to vary due to randomly dropped neurons. To estimate uncertainty, the detector is run $K$ times on the same PPG signal $x_i$, producing $K$ predictions for the AF class:
\begin{equation}
\{\hat{p}_1(x_i), \hat{p}_2(x_i), \dots, \hat{p}_K(x_i)\},
\end{equation}
from which the mean prediction is computed as
\begin{equation}
\bar{p}(x_i) = \frac{1}{K} \sum_{k=1}^K \hat{p}_k(x_i).
\end{equation}
Uncertainty is quantified using the predictive entropy
\begin{equation}
\mathcal{H}(x_i) = -\bar{p}(x_i)\ln(\bar{p}(x_i) + \epsilon) - (1 - \bar{p}(x_i))\ln(1 - \bar{p}(x_i) + \epsilon),
\end{equation}
where $\epsilon$ prevents numerical instability for $\ln(0)$. The entropy is bounded between 0 and $\ln(2)$, with larger values indicating higher uncertainty. Overall uncertainty is reported as the average entropy across all PPG signals:
\begin{equation}
\mathcal{\bar{H}} = \frac{1}{N} \sum_{i=1}^N \mathcal{H}(x_i).
\end{equation}
During testing on the investigation dataset, $K = 50$, the dropout rate is set to 0.5, and $\epsilon = 10^{-10}$.

\backmatter

\bmhead{Acknowledgements}
This work was supported by the Research Council of Lithuania [grant number S-MIP-24-73] and by the European Partnership on Metrology [grant number 22HLT01].

\section*{Declarations}

\bmhead{Funding}
This work was supported by the Research Council of Lithuania [grant number S-MIP-24-73] and by the European Partnership on Metrology [grant number 22HLT01].

\bmhead{Conflict of interest/Competing interests}
The authors declare that they have no known competing financial interests or personal relationships that could have influenced the work reported in this paper.

\bmhead{Ethics approval and consent to participate}
The data acquisition protocol was approved by the Kaunas Region Biomedical Research Ethics Committee (No. BE-2-20).

\bmhead{Author contribution}
Andrius Solo\v{s}enko: Data curation, Software, Formal analysis, Investigation, Visualization, Writing -- review \& editing. Enik\"o Vargova: Formal analysis, Investigation. Lo\"ic Coquelin: Formal analysis. Vaidotas Marozas: Writing -- review \& editing, Funding acquisition. Andrius Petr{\.{e}}nas: Conceptualization, Methodology, Writing -- original draft.

\bibliography{bibUnct}

\end{document}